\documentclass[preprint,showpacs,preprintnumbers,amsmath,amssymb,numerical]{revtex4}

\usepackage{graphicx}
\usepackage{dcolumn}
\usepackage{bm}
\usepackage{color}
\graphicspath{{C:/Users/Marco/Documents/Resonator_project/APL_manuscript_2014/figures/}}
\begin{document}


\title{Microwave dielectric losses at millikelvin temperatures of substrates suitable for High critical Temperature Superconductors epitaxial growth}


\author{M. Arzeo$^1$}

\author{F. Lombardi$^1$}

\author{T. Bauch$^1$}

\affiliation{$^1$ Quantum Device Physics Laboratory, Department of Microtechnology and Nanoscience, Chalmers University of Technology,  SE-41296 G\"{o}teborg, Sweden}

\date{\today}

\begin{abstract}
We have investigated both the temperature and the power dependence of microwave losses for various dielectrics commonly used as substrates for the growth of High critical Temperature Superconductor thin films. We present measurement of niobium superconducting $\lambda/2$ coplanar waveguide resonators, fabricated on $La_{0.3}Sr_{0.7}Al_{0.65}Ta_{0.35}O_3$ (LSAT), $MgO$ (MgO) and $LaAlO_3$ (LAO), at the millikelvin temperature range and at low input power. By comparing our results with the Two-Level System (TLS) model, we have discriminated among different dominant loss mechanisms. LSAT has shown the best results as regards the dielectric losses in the investigated regimes. 
\end{abstract}

\maketitle

The study of superconducting circuits operating in the microwave regime has attracted great attention during the past decade. The extremely low conductor losses makes superconductors very attractive for a vast range of microwave applications such as narrow bandpass filters\cite{Simon}, energy resolved photon detectors\cite{Day}, circuit quantum electrodynamics\cite{wallraff} and quantum bits\cite{Devoret}. Dielectric losses in the substrate material have been recently identified as one of the main sources of decoherence in superconducting quantum circuits, limiting the ultimate performance of these devices. These losses, which are very pronounced at mK temperatures and low microwave powers (single photon limit), representing the typical operation condition of quantum circuits, have been attributed to a bath of quantum two level systems (TLS) in the dielectric substrate material\cite{OConnel,Lindstrom,Macha,Sage,Wang,Gao}. 
Extensive studies on microwave losses, related to TLS in the dielectric substrate,  have been mainly preformed on sapphire and $SiO_2/Si$, which are the materials  commonly used for low critical temperature superconductor (LTS) circuits.  At present, substrate materials compatible with the epitaxial growth of High critical Temperature Superconductors (HTS) have not yet been characterized in the low temperature (mK) and low power (few photons) limit. This is of fundamental importance for the development of  LTS/HTS hybrid  devices which could overcome the  state of art performances of  quantum devices possibly introducing new functionalities connected with the d-wave symmetry of the order parameter in HTS\cite{lucignano}. 
At the same time HTS based quantum devices operating at mK temperature  would allow the measurement of the low energy quasiparticle spectrum with unprecedented energy resolution\cite{Thilo, Gustafsson}. This is thought to be fundamental for understanding the mechanism leading to high critical temperature superconductivity, still unsettled after more than 25 years from their discovery. 

In this letter we compare the microwave properties, as a function of temperature and microwave power (number of photons), of various dielectric materials commonly used as substrates for the epitaxial growth of HTS thin films. The aim is to find the most suitable substrates for HTS microwave applications in the low power and temperature limit. 

We report on both temperature and power dependence of microwave dielectric losses for $MgO$ ($\epsilon_r\sim9$), $LaAlO_3$ (LAO, $\epsilon_r\sim25$) and $La_{0.3}Sr_{0.7}Al_{0.65}Ta_{0.35}O_3$ (LSAT, $\epsilon_r\sim22$); here $\epsilon_r$ is the corresponding relative permittivity at 300 K. We have compared our data with the Two-Level System (TLS) resonant absorption theory\cite{Strom} to discriminate among the possible loss mechanisms. 
A common way to measure the microwave losses of a dielectric material is to measure the internal quality factor of a superconducting resonator patterned on a dielectric substrate\cite{OConnel}. When the resonator is coupled to the external environment, the total resonator quality factor $Q_{tot}$ is the parallel of the external $Q_{ext}$ and the internal $Q_{int}$ ones: $Q_{tot}= [Q_{ext}^{-1} + Q_{int}^{-1}]^{-1}$. The first is determined by the coupling to the measurement setup, the second, instead, is given by the sum over all the loss mechanisms (e.g. dielectric, conductor, radiation losses, etc) intrinsically affecting the device. $Q_{int}$ is generally expressed by the loss tangent: $\tan{\delta}=Q^{-1}_{int}=\sum_{i} Q^{-1}_{i}$. In particular, microwave dielectric losses ($\tan{\delta_d}$) are proportional to the sum over all the possible microwave absorption mechanisms ($\alpha_n$): $\tan{\delta}_{d}\propto\sum_{n}\alpha_{n}$. 

Microwave losses in a superconducting resonator depend both on temperature and probing power.\cite{Lindstrom,Macha,Sage}  For temperatures above $T^*\sim\frac{T_c}{10}$, with $T_c$ the superconducting transition temperature, conductor losses of LTS, described by the \emph{Mattis-Bardeen} theory\cite{Mattis}, increase with temperature due to quasiparticles generation. However for temperatures below $T^*$ ($T^*\simeq 1$ K for Nb) conductor losses saturate and other microwave losses such as dielectric losses, $\tan{\delta_d}$, become dominant and one can generally write for the measured loss tangent:
\begin{equation}
\tan{\delta} (T,P)=\tan{\delta_d} (T,P) + \tan{\delta_L}, 
\label{eq:tgt}
\end{equation}
where $\tan{\delta_L}$ is a background term (including conductor, radiation and other loss mechanisms), which we treat as constant in the studied temperature and applied power range. Dielectric losses can be modeled by assuming the presence of a bath of TLS whose electric dipole moment, $\vec{d}$, is coupled to the electric field in the resonator. In the case of resonant absorption due to TLS with an energy splitting of $h f_0$, the temperature and power dependence of the dielectric losses can be derived:\cite{Strom}
\begin{equation}
\tan{\delta}_{d} (T,P)= F\alpha_{TLS}  \left(1+\frac{P}{P_c}\right)^{\!-1/2} \tanh{\left(\frac{h f_0}{2 k_B T}\right)},
\label{tgd}
\end{equation}
 with $\alpha_{TLS}=\frac{\pi N d^{2}}{3\epsilon}$. $F$ is the filling factor, which depends on the geometry and the electric field distribution, $N$ is the density of states of the TLS, $\epsilon=\epsilon_0\epsilon_r$ is the absolute permittivity of the host material and $\epsilon_0$ the vacuum permittivity. At fixed temperature, $\tan{\delta}_{d}$ saturates for power $P$ below the critical value $P_c$ (low power limit): $\tan{\delta}_{d} (T)= F\alpha_{TLS}  \tanh{\left(h f_0/2 k_B T\right)}$. By increasing the power above $P_c$, the losses decrease as $P^{-1/2}$. In the high power limit the bath of TLS is then saturated by the large number of photons of energy $hf_0$ circulating in the device. The critical power is given by: $P_c=3\hbar^2 \epsilon/2d^2T_1T_2$, where $T_1$ and $T_2$ are, respectively, the relaxation and the dephasing times of the TLS in the specific dielectric materials. The product $F\alpha_{TLS}$ corresponds to the value of the ultimate dielectric losses, $\tan{\delta_d}$, in the low-temperature ($k_B T\ll hf_0/2$) and low-power limit ($P\ll P_c$). 

To model properly the power dependence, a modification of Eq. \ref{tgd} is required since not all the microwave power at the input $P_{inc}$ is transfered into the device. It is, in fact, more convenient to introduce the circulating power $P_{cir}$, defined as: $P_{cir}= P_{inc}\left[\frac{Q^2_{int} Q_{ext}}{\pi(Q_{int}+Q_{ext})^2}\right]$ and which is a measure of the average number of photons in the resonator\cite{Sage}: $N_{ph}=P_{cir}/hf^2_0$. Since the microwave field is not uniform over the volume $V_f$ containing the bath of TLS, one can not simply replace $P$ with $P_{cir}$ in the Eq. (\ref{tgd}). A full theoretical treatment would require the numerical calculation of the spatial field distribution over $V_f$. However, it is possible (as already shown in refs.\cite{Wang,Sage}) to fit the data to a modification of  Eq. (\ref{tgd}):
\begin{equation}
\tan{\delta}_{d} (P_{cir})= F\alpha_{TLS}  \left[1+\left(\frac{P_{cir}}{P'_c}\right)^{\beta}\right]^{-1/2} \tanh{\left(\frac{h f_0}{2 k_B T}\right)}.
\label{tgd2}
\end{equation}  
Here, $\beta$ is an additional fitting parameter and $P'_c$ is the critical power depending also on the geometry of the resonator. 

The resonator fundamental resonance frequency in the low power limit is also dependent on temperature. Applying the \emph{Kramers-Kronig} relations to Eq. (\ref{tgd})\cite{Strom}, the contribution from the resonant absorption of TLS to the variation in temperature of the resonance frequency, defined as $\delta f_0 (T)= \frac{f_0 (T) - f_0 (T_0)}{f_0 (T_0)}$, can be expressed as follows:\cite{Strom}
\begin{equation}
\delta f_0 (T)=\frac{F\alpha_{TLS}}{\pi}\left\{\ln{\left(\frac{T}{T_0}\right)}-[g(T,f_0) - g(T_0,f_0)]\right\},
\label{eq3}
\end{equation}
where $g(T,f)=Re[\Psi(1/2+ h f/2\pi i k_B T)]$, $\Psi$ is the complex digamma function and $T_0$ is a reference temperature. The digamma function is significant only for temperatures such that $k_B T \leq hf_0/2$ and it results in an upward turn below $T=hf_0/2k_B$. In our specific case ($f_0\simeq5$ GHz), it corresponds to $T\simeq120$ mK which we set as reference temperature $T_0$ for our data.  

\begin{figure}[b]
\begin{center}
 \includegraphics[scale=0.3]{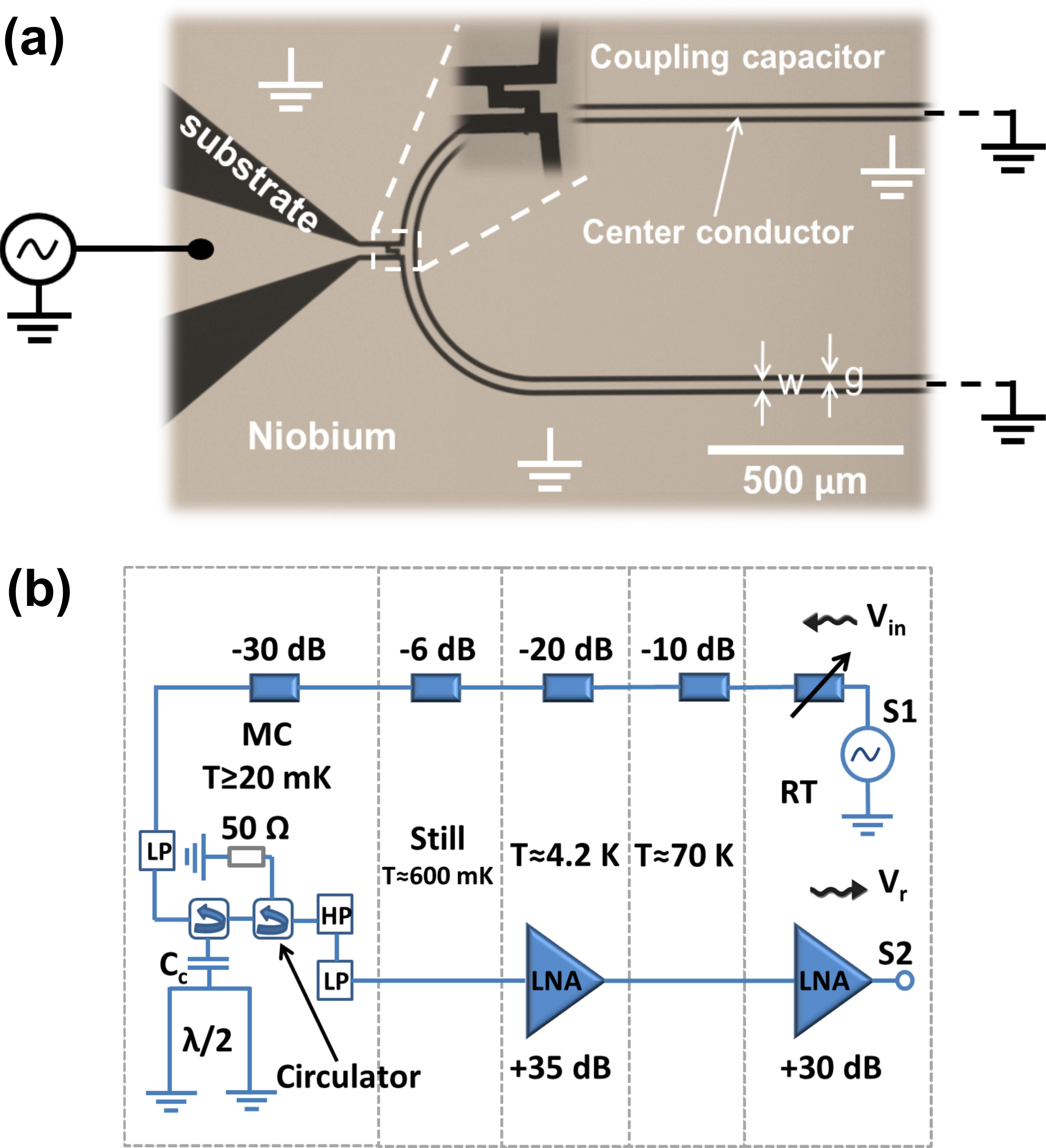}
 \caption{Resonator geometry and measurement setup. (a) Optical image of our device. In the image, the main parts of the resonator are highlighted. A zoom in shows the finger coupling capacitance to the input/output feedline. (b) Schematic of the measurement setup; MC: Mixing Chamber, LP and HP: low- and high-pass filter, LNA: low noise amplifier. In this configuration $S_{21}=V_r/V_{in}$ is a measurement of the complex reflected signal from the device.  \label{Figure1}}
\end{center}
 \end{figure}

In this work, we present measurements performed on half-wavelength ($\lambda/2$) superconducting coplanar waveguide (CPW) resonators whose geometry is shown in Fig.\ref{Figure1}(a). A $200$ nm thick Nb film is directly deposited on polished and annealed substrates\cite{Megrant} by magnetron sputtering in ultra high vacuum ($p\leq10^{-8}$ mbar).  The film is then patterned by reactive ion etching (RIE) with $NF_3$ plasma through an e-beam defined resist mask. The geometry of the resonator (see Fig. \ref{Figure1}(a)) is chosen such to minimize radiation losses and cross talk between adjacent lines. 
Since microwave dielectric losses induced by TLS are geometry dependent,\cite{Gao, Sage}a direct comparison between different dielectrics requires the use of resonators of identical geometry. Here the value of the center conductor width and its separation gap to the ground plane are respectively $w=20 \;\mu$m and $g=12 \;\mu$m (see Fig. \ref{Figure1}(a)).  
The resonators lengths are, instead, chosen such that the fundamental resonance frequency ($f_0$) is around $5$ GHz. Each CPW resonator is capacitively coupled to the input/output feedline with a finger capacitance (see zoom in Fig.\ref{Figure1}(a)) at the voltage anti-node for the standing electromagnetic wave, while the two ends are shorted to the ground plane. The device chips are mounted in a rf-tight copper box which is thermally anchored to the mixing chamber of a dilution (dry) fridge. 

A multilayer $\mu$-metal shield is used to screen the ambient magnetic field. The microwave reflectivity signal $S_{21}=V_r/V_{in}$ from the resonators  (see Fig. \ref{Figure1}(b))  has been measured as a function of temperature and input power. The attenuation at the input line (see Fig. \ref{Figure1}(b)) allows to probe the devices at power values corresponding to a few photons populating the resonator. The reflected signal of the resonator is amplified by using a cryogenic amplifier at the 4 K stage and a room temperature amplifier. The quality factors, together with the fundamental resonance frequency $f_0$, have been determined by fitting the complex valued reflection signal $S_{21}$ around $f_0$ to the expression:
\begin{equation}
\Gamma=\frac{Q_{ext}^{-1} - Q_{int}^{-1} -2i\delta x}{Q_{ext}^{-1} + Q_{int}^{-1} +2i\delta x},
\label{eq:gamma}
\end{equation}
where $\delta x=\frac{f-f_0}{f_0}$. 

\begin{figure} [b]
\begin{center}
 \includegraphics[scale=0.3]{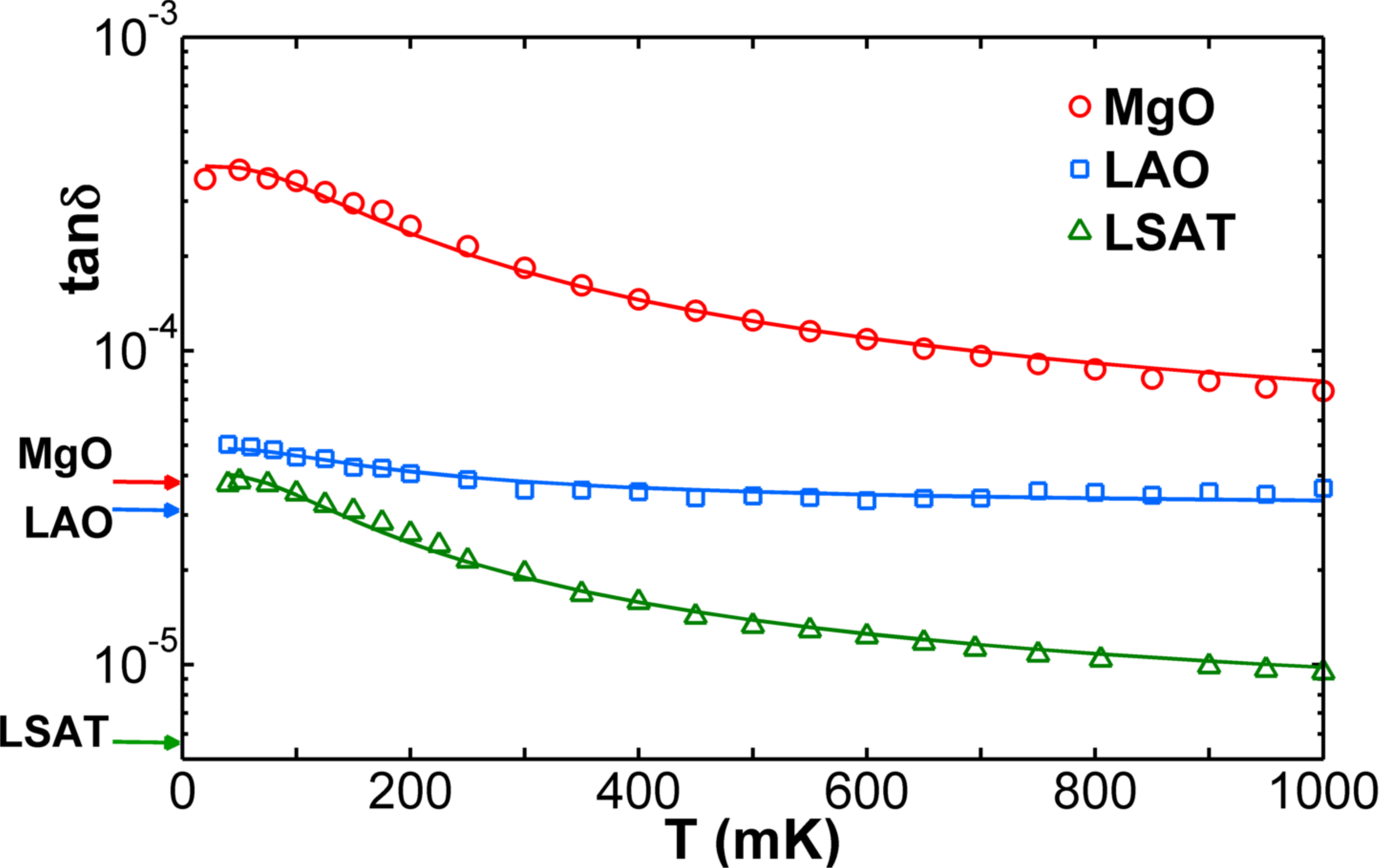}
 \caption{Temperature dependence in the low power limit of the resonator internal losses for all the studied substrates. Solid lines represent the best fit to Eq. (\ref{tgd}). The arrows indicate the value of the background loss, $\tan{\delta_L}$. \label{Figure2}}
\end{center}
 \end{figure}

In Fig. \ref{Figure2} we show the measured loss tangent as a function of bath temperature in the low power limit for three different dielectric substrates. By fitting the obtained experimental data to Eq. \ref{eq:tgt} and  Eq. \ref{tgd} (see Fig. \ref{Figure2}), neglecting the power dependence, we have extracted the values of $F\alpha_{TLS}$ and $\tan{\delta_L}$ for all the investigated dielectric materials. The obtained results are listed in Table \ref{Table1}. 

For MgO and LSAT we observe a strong temperature dependence with a monotonic decrease of the loss tangent with increasing temperature, in good agreement with the TLS model. Indeed, the large ratio $F\alpha_{TLS}/\tan{\delta_L}>1$ (see Table  \ref{Table1}) suggests that resonant absorption by TLS is dominating over other loss mechanisms included in the background term. The background loss $\tan{\delta_{L}}$ obtained from the fit of the temperature dependence is most probably determined by the onset of microwave losses caused by relaxation absorption\cite{Hunk} at high temperatures. In the case of the CPW resonator on the LAO substrate, instead, an almost flat temperature dependence has been observed. Indeed, the small value of the ratio $F\alpha_{TLS}/\tan{\delta_L}$ is a strong indication that the background losses, $\tan{\delta_L}$, are dominating in the investigated temperature range. We attribute the large value of $\tan{\delta_L}\sim3\times 10^{-5}$, which is roughly an order of magnitude larger than the high power limit of bare LAO at 4K\cite{zuccaro}, to the conductor losses of Nb. The relatively high conductor losses of Nb, in this case, might be related to inhomogeneities in the film that are caused by twin domains in the LAO. Detrimental effects on the microwave properties induced by twin domains in YBCO films grown on LAO are well known and have been extensively studied by low temperature laser scanning microscopy\cite{twin}.

\begin{table}[t]
\caption{$F\alpha_{TLS}$ and $\tan{\delta_L}$ values for the different dielectrics, extracted from fits to the equations indicated in parenthesis. The values we obtained using Sapphire as substrate for our devices are reported as reference and comparison with data reported in literature. \cite{Gao, Lindstrom, Macha,Sage} \label{Table1} }
 \begin{tabular}{c c c c c}
\hline
\hline
\textbf {Substrate}\hspace{2mm} & MgO \hspace{2mm} & LAO \hspace{2mm} &  LSAT  \hspace{2mm}& Sapphire\\
\hline
\textbf{$F\alpha_{TLS}$} (1,2) & $3.3\times10^{-4}$ & $1.8\times10^{-5}$& $3.4\times10^{-5}$ & $7.9\times10^{-6}$\\ 
\textbf{$F\alpha_{TLS}$} (3)& $5.5\times10^{-4}$& N.A. & $4.6\times10^{-5}$&$4.8\times10^{-6}$ \\ 
\textbf{$F\alpha_{TLS}$} (4)& $4.9\times10^{-4}$& $2.5\times10^{-5}$& $4.6\times10^{-5}$& $1.4\times10^{-5}$\\ 
\textbf{$tan\delta_{L}$} (1,2)& $3.8\times10^{-5}$& $3.1\times10^{-5}$& $5.6\times10^{-6}$& $3.6\times10^{-6}$\\ 
$\frac{F\alpha_{TLS}}{tan\delta_{L}}$ (1,2)&8.7 &0.6 &6.1 &2.2\\
\hline
\hline 
 \end{tabular}
\end{table}

\begin{figure} [t]
\begin{center}
 \includegraphics[scale=0.3]{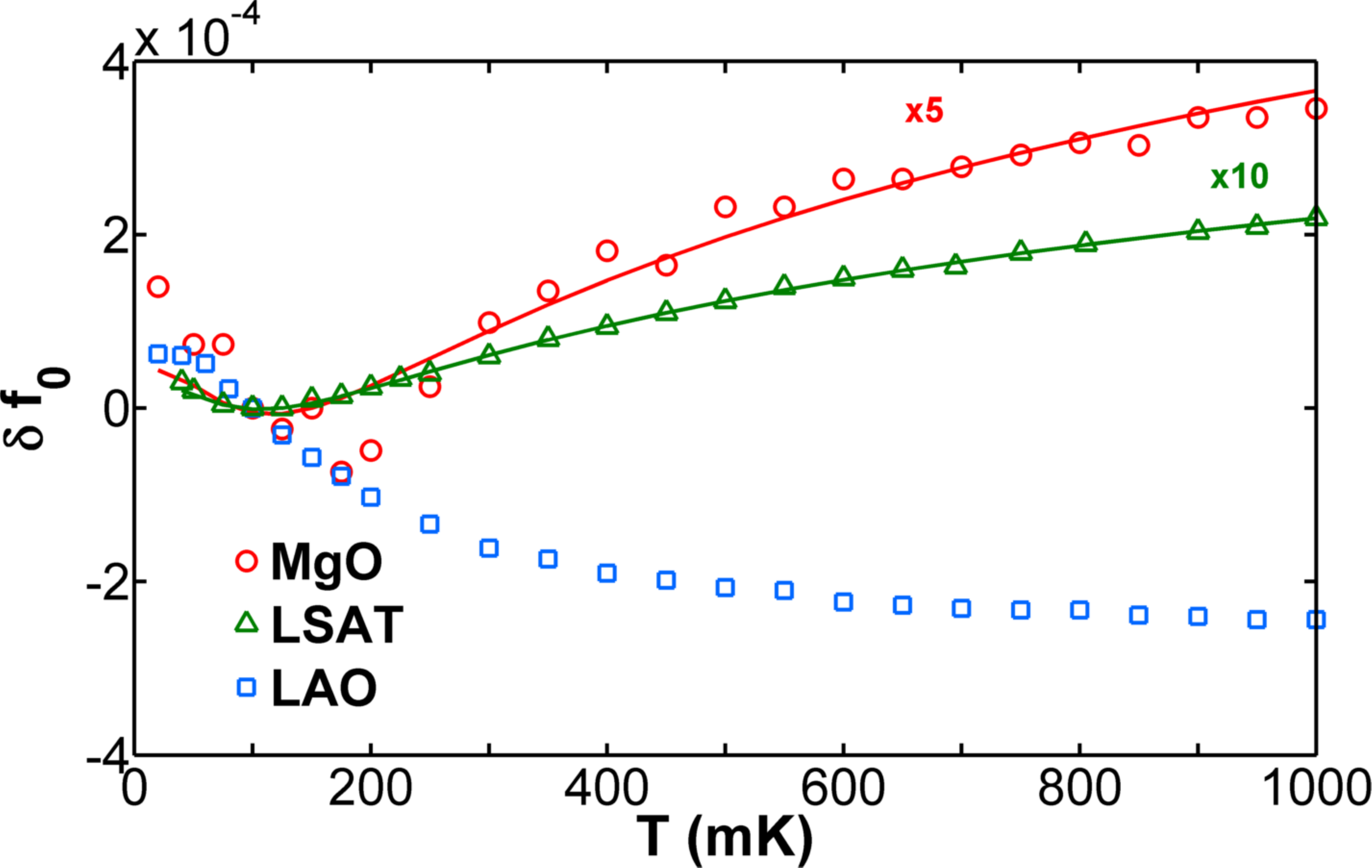}
 \caption{Temperature dependence in the low power limit of the resonance frequency variation for all the studied substrates. Solid lines represent the best fit to Eq. (\ref{eq3}). Data for MgO and LSAT are, respectively, multiplied by a factor 5 and 10, in order to get a clearer visualization. \label{deltaf}}
\end{center}
 \end{figure}

The measured resonance frequency shift $\delta f_0$ as a function of the bath temperature in the low power limit for the studied dielectrics are shown in Fig. \ref{deltaf}. By fitting data of $\delta f_0$  to the Eq. (\ref{eq3}), the value of $F\alpha_{TLS}$ can be obtained as well. Equation (\ref{eq3}) accounts only for dielectric losses due to resonant absorption from the TLS bath. A comparison with our data allows to further discriminate among the possible loss mechanisms. 

The temperature dependence of $\delta f_0$ relative to LSAT follows very well the theoretical prediction of Eq. (\ref{eq3}) with the characteristic upturn of $\delta f_0$ for temperatures smaller than $T_0= hf_0/2k_B$. Moreover, the extracted value of $F\alpha_{TLS}$ is in good agreement with the one determined from the temperature dependence of $\tan{\delta}$ (see Table \ref{Table1}). For MgO, the fit of $\delta f_0 (T)$ is rather good with some slight departure of the experimental data from the theoretical prediction in the low temperature regime $T<T_0$. However, the value of $F\alpha_{TLS}$ extracted for $T>T_0$ is consistent with the one obtained from the loss tangent data (see Table \ref{Table1}). In the case of LAO, instead, we observe a monotonic decrease in temperature of $\delta f_0$ , which does not allow any determination of the $F\alpha_{TLS}$ value. This anomalous behavior supports the fact that the twinned LAO substrate induces inhomogeneities in the Nb film strongly affecting its microwave properties, which resembles the variation of an effective kinetic inductance as a function of temperature\cite{Gao2}.

\begin{figure} [h!]
\begin{center}
 \includegraphics[scale=0.3]{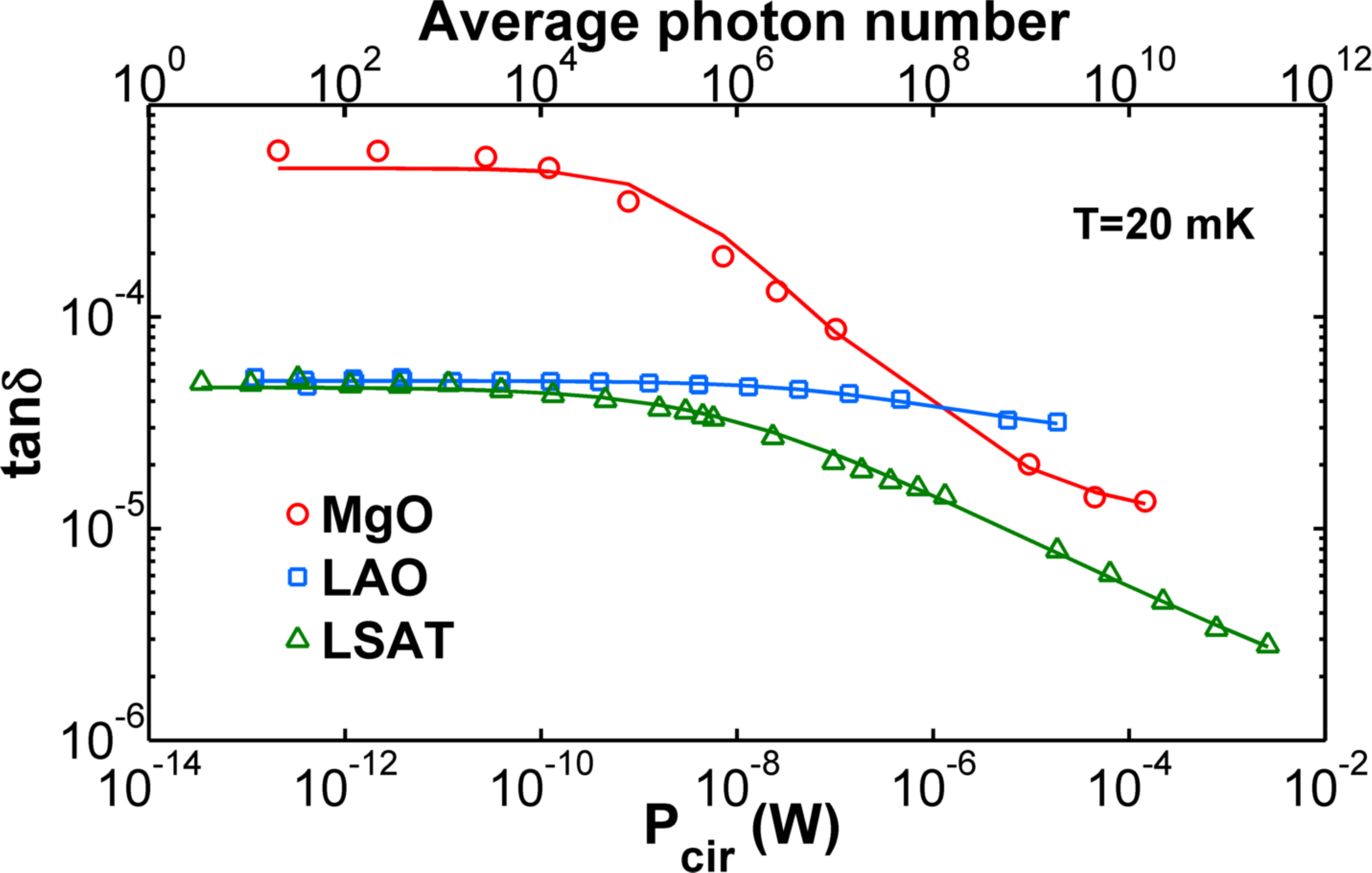}
 \caption{Microwave losses as a function of the circulating power and the average number of photons in the resonator, for all the studied substrates. Solid lines represent the best fit to the sum of Eq. (\ref{tgd2}) and an additional background term. \label{fig:power} }
\end{center}
 \end{figure}

Figure \ref{fig:power} shows the measured loss tangent as a function of the power (bottom axis) and the average number of photons (top axis) circulating in the resonator at 20 mK for the three dielectrics. The values $F\alpha_{TLS}$ extracted from the fit of the power dependence are consistent with the previous analyses (see Table \ref{Table1}). Our data on LAO show almost no power dependence for the relative microwave losses, supporting once more the assumption that microwave losses are dominated by those from the superconductor. For both MgO and LSAT, instead, we observe a strong reduction of $\tan{\delta}$ by increasing the power above the critical value, in agreement with a progressive saturation of the TLS bath (see Fig. \ref{fig:power}). Due to limitation in the measurement setup, optimized to work at low input power, it was not possible to reach the saturation of $\tan{\delta}$ at high power and to properly extract the background term.

In conclusion, we have reported on the temperature and power dependence of microwave dielectric losses and resonance frequency shift of CPW resonator patterned on three different dielectric substrates suitable for the epitaxial growth of HTS thin films. The different analyses are consistent and show clear evidences about microwave loss mechanisms for the three investigated dielectrics. Our results indicate LSAT as an excellent choice for substrate for HTS or, in general, complex oxides based microwave quantum devices. We have shown that its intrinsic dielectric losses, in the millikelvin temperature range, are dominated by TLS resonant absorption. This will help future device designs and fabrication optimization to minimize the effect of this loss mechanism, since geometry dependence and the origin of TLS in such devices is well known at the modern state of art.\cite{Gao,Wang,Sage,Megrant} Dielectric losses in MgO are also quite well represented by the TLS resonant absorption model. However the relative large value of $F\alpha_{TLS}$ makes MgO not very suitable for microwave applications in the mK and single photon operation regime. This has also very important consequences for the employment of MgO as tunnel barrier in spintronic related quantum devices and epitaxial NbN films grown on MgO for kinetic inductance based devices such as single photon detectors. In the case of LAO, in contrast to the other dielectric substrates, we have observed a very weak temperature and power dependence of the loss tangent $\tan{\delta}$ and a monotonic decrease of the resonance frequency shift $\delta f_0$ in the millikelvin range. Therefore the observed losses are most probably related to relatively higher conductor losses in the Nb film due to the presence of inhomogeneities in the Nb film induced by the twin domains in the LAO.         

\vspace{1cm}
This work has been partially supported by the Swedish Research Council (VR) and the Knut and Alice Wallenberg Foundation. We acknowledge support from the Marie Curie Initial Training Action (ITN) Q-NET 264034.



\end{document}